\documentclass{article}

\usepackage{arxiv}

\usepackage{graphicx}
\usepackage[utf8]{inputenc} % allow utf-8 input
\usepackage[T1]{fontenc}    % use 8-bit T1 fonts
\usepackage{hyperref}       % hyperlinks
\usepackage{url}            % simple URL typesetting
\usepackage{booktabs}       % professional-quality tables
\usepackage{amsfonts}       % blackboard math symbols
\usepackage{nicefrac}       % compact symbols for 1/2, etc.
\usepackage{microtype}      % microtypography
\usepackage{lipsum}

\usepackage{tikz}
\usepackage[square,numbers]{natbib}
\usepackage{algpseudocode,algorithm,algorithmicx} 
\usepackage{enumerate}
\usepackage{amsmath}
\usepackage{float}

\title{Multiple Change Point Detection and Validation in Autoregressive Time Series Data}

\author{
  Lijing ~Ma \\
  Department of Mathematics and Statistics \\
  Macquarie University \\
  \texttt{lijing.ma@mq.edu.au} \\
   \And
  Andrew ~Grant \\
  MRC Biostatistics Unit \\
  University of Cambridge \\
  \texttt{andrew.grant@mrc-bsu.cam.ac.uk} \\
   \And
  Georgy ~Sofronov \\
  Department of Mathematics and Statistics \\
  Macquarie University \\
  \texttt{georgy.sofronov@mq.edu.au} \\
}

\begin{document}
\maketitle

\begin{abstract}
It is quite common that the structure of a time series changes abruptly. Identifying these change points and describing the model structure in the segments between these change points is of interest. In this paper, time series data is modelled assuming each segment is an autoregressive time series with possibly different autoregressive parameters. This is achieved using two main steps. The first step is to use a likelihood ratio scan based estimation technique to identify these potential change points to segment the time series. Once these potential change points are identified, modified parametric spectral discrimination tests are used to validate the proposed segments. A numerical study is conducted to demonstrate the performance of the proposed method across various scenarios and compared against other contemporary techniques.
\end{abstract}

% keywords can be removed
\keywords{Changepoint detection \and Autoregressive time series \and Likelihood ratio scan statistics \and Multiple testing problems}

\section{Introduction}
\label{sec:intro}
The statistical properties of time series data, such as mean and variance or the coefficients of the regression model, may change abruptly at unknown time points. Identifying those unknown time points is referred to as change point detection or time series segmentation. The change point problem was first considered by \cite{page1954continuous} and \cite{page1955test} for quality control. Since then, the topic has been explored theoretically and computationally in the field of statistics and computer science, and has been applied to economics \cite{bai2003computation} \cite{bai2010common}, finance \cite{aue2013structural} \cite{andreou2009structural}, and biology \cite{olshen2004circular} \cite{niu2012screening}. Furthermore, see the recent survey papers by \cite{jandhyala2013inference}, \cite{Aminikhanghahi2017} and \cite{TRUONG2020107299} for the development of univariate or multivariate time series segmentation methods.

There are essentially two types of approaches for detecting unknown change points under parametric design: the model selection method and the traditional hypothesis testing method. Model selection or exact segmentation methods generally include two elements, a cost function and an optimization algorithm. The computational complexity depends on the complexity of data and the number of change points. In contrast, the approximate segmentation methods have significantly less computational cost when there are more change points. Here, we follow in the direction of the approximate segmentation methods. 

One popular representative of the approximate segmentation methods is the binary segmentation (BS) family of methods. The core idea is that BS tests if there is a change point in the process at each step or iteration (see \cite{fryzlewicz2014wild} for a detailed description). BS has gained huge popularity due to the minor computational cost and its user-friendliness. However, the method may ignore change points if the length of the segment is relatively short. Hence, \cite{olshen2004circular} further improved the BS algorithm, and proposed the circular BS (CBS) method. \cite{fryzlewicz2014wild} proposed the wild BS (WBS) approach to detect the number and locations of changes in a piecewise stationary model when the values of the parameters change. Another representative of the approximate segmentation methods is bottom-up segmentation, which is less explored than the BS algorithm (we recommend the paper by \cite{keogh2001online} for further details). Bottom-up segmentation is also easy to apply: the first step is to obtain a sequence of overestimated change points; the second step is to eliminate the falsely-detected ones.

However, both the BS algorithm and the bottom-up method may suffer from the multiple testing problem. \cite{eichinger2018mosum} mentions in regards to the BS algorithm that "it can be difficult to interpret the results in terms of significance due to the multiple testing involved".
Thus, \cite{fryzlewicz2014wild} added a randomized segment selection step to the BS method. \cite{li2016fdr} proposed multiscale change point segmentation with controlled false discovery rate (FDR) based on multiscale statistics considered by \cite{frick2014multiscale} for inferring the changes in the mean of an independent sequence of random variables. \cite{cao2015changepoint} developed a large scale multiple testing procedure for data with clustered signals. The earlier references that introduced FDR for multiple change point detection include \cite{niu2012screening} and \cite{hao2013multiple}, which are motivated by genome data. Hitherto, only a small amount of literature attempts to address this issue. When the observations are dependent, detecting multiple change points is quite a difficult task, especially in the case of autoregressive processes. \cite{davis1995testing} studied the asymptotic behavior of the likelihood ratio statistic in testing if a change point has occurred in the mean, the autocovariance structure or the order of an autoregressive process. Later on, \cite{davis2006structural} estimated all the parameters of a piecewise stationary autoregressive process by using a genetic algorithm to optimize an information criterion as objective function. \cite{chakar2017robust} proposed a robust approach for estimating change points in the mean of an AR(1) process.  \cite{korkas2017multiple} upgraded the WBS algorithm by applying a locally stationary wavelet process for estimating change points in the second-order structure of a piecewise stationary time series model. \cite{yau2016inference} proposed a likelihood ratio scan algorithm (LRSM) to estimate change points in piecewise stationary processes.

In this paper, we develop a new Multiple Comparisons Procedure for a Multiple change point Problem (MCP-MCP, or MCP2 for short), to estimate the number and locations of change points in a piecewise stationary autoregressive model. The procedure includes three simple steps: the first step is to apply the likelihood ratio scan statistics by \cite{yau2016inference} to obtain a set of potentially overestimated change points; the second step is to use the spectral discrimination procedure developed by \cite{grant2017parametric} to eliminate possibly falsely discovered change points; the third step is to use a classic controlling FDR procedure and an adjusted p-value Bonferroni procedure to address the multiple testing issue. Our work is mainly inspired by \cite{yau2016inference} and \cite{korkas2017multiple} and, to the best of our knowledge, is the first paper to address the multiple testing issue taking the dependency into account as a bottom-up segmentation method.  

As indicated by \cite{mercurio2004statistical}, it is highly risky to treat non-stationary data as though they are from a stationary process when making predictions and forecasting. Therefore, the estimation accuracy tends to be very important and the exact properties of estimates needs careful attention. In our simulation study, we focus on the correct estimated number and locations of change points. The simple but meaningful plots we present are used to show the robustness of estimates for each method. The structure of the paper is as follows. In section \ref{sec:methods1}, we provide the details of the MCP2 method. In section \ref{sec:sims}, through extensive simulation experiments and in section \ref{sec:data}, through two real data examples, we evaluate the performance of the MCP2, LRSM and WBS methods. Lastly, we conclude the paper in section \ref{sec:discussion} with discussion and comments on future research.

\section{A Multiple Comparisons Procedure for Change Point Detection}
\subsection{Non-stationary time series segmentation as a multiple testing problem}
\label{sec:methods1}
We start this section by demonstrating the autoregression process segmentation problem, and how it can be viewed as a multiple hypothesis testing problem. Let $x_{1}, x_{2},\dots,x_{T}$ be a sequence of an autoregression process, with $q$ the unknown number of change points and $k_{1}, k_{2},\dots, k_{q}$ their respective unknown positions, where $1 < k_{1} < k_{2} < \cdots < k_{q} < T$. The autoregression process with multiple change points is illustrated as below
\begin{equation} \label{2.1} 
x_{t}=
\begin{cases}
\beta_{0}^{(1)} + \beta_{1}^{(1)} x_{t-1} + \cdots + \beta_{p_{1}}^{(1)} x_{t-p_{1}} + \varepsilon_{t}^{(1)}, &t=1,\dots,k_{1}\\
\beta_{0}^{(2)} + \beta_{1}^{(2)}x_{t-1} + \cdots + \beta_{p_{2}}^{(2)} x_{t-p_{2}} + \varepsilon_{t}^{(2)}, &t=k_{1}+1,\dots,k_{2}\\
\cdots, \cdots \\
\beta_{0}^{(q+1)} + \beta_{1}^{(q+1)}x_{t-1} + \cdots + \beta_{p_{q+1}}^{(q+1)} x_{t-p_{q+1}} + \varepsilon_{t}^{(q+1)}, &t=k_{q}+1,\dots,T\\
\end{cases}
\end{equation} 
where $\varepsilon_{t} \sim i.i.d.N(0,\sigma^2_{t})$ and each segment is a stationary autoregression of order $p$ (AR(p)) and independent of each other. This problem can be expressed as a classical single hypothesis testing problem, as follows. Letting $\theta_{t}$ be the parameters that generate the data at each time point, $t = 1, \dots, k_{q}, \dots,T$,
\begin{eqnarray} \label{2.2}
H_{0} &:& \theta_{1} =  \cdots = \theta_{k_{q}+1}  = \cdots = \theta_{T} \nonumber\\
H_{1} &:& \theta_{1} = \cdots = \theta_{k_{1}}  \neq \theta_{k_{1}+1} = \cdots = \theta_{k_{2}} \neq \cdots \neq \theta_{k_{q}+1} = \cdots = \theta_{T}
\end{eqnarray}
%\end{equation}            
%versus the alternative hypothesis: 
%\begin{equation}
If $H_{1}$ is supported, the data are split into $ q+1 $ segments, ($x_{1}, x_{2}, \dots, x_{k_{1}}$), ($ x_{k_{1}+1}, x_{k_{1}+2}, \dots, x_{k_{2}} $), $\dots$, ($ x_{k_{q}+1}, x_{k_{q}+2}, \dots, %x_{k_{q+1}} $),
x_{T} $),
with different generating parameters for each segment denoted by $\theta_{i} : = (p_{i},\beta_{p_{i}}^{(i)},{\sigma^2}^{(i)})$, $i = 1, \dots, q+1$. 

The ambitious objective is to estimate the number of change points $q$, the location vector $ k = (k_{1}, k_{2}, \cdots, k_{q}) $ and the parameters for each segment $\theta_{i}$. It is not practical to achieve this objective through the aforementioned single hypothesis testing framework, hence we decompose (\ref{2.2}) to multiple hypothesis tests              
\begin{eqnarray} \label{2.3}
H_{0}(i) &:& \theta_{k_{i-1}+1 : k_{i}} = \theta_{k_{i}+1 : k_{i+1}}  \nonumber\\
H_{1}(i) &:& \theta_{k_{i-1}+1 : k_{i}} \neq \theta_{k_{i}+1 : k_{i+1}}
\end{eqnarray}
for $i = 1, \ldots, q$. Since we assume that each segment is an independent time series, (\ref{2.3}) can be viewed as a multiple testing problem by determining whether two adjacent segments ($ x_{k_{i-1}+1} , x_{k_{i-1}+2}, \dots, x_{k_{i}} $) and ($ x_{k_{i}+1}, x_{k_{i}+2}, \dots, x_{k_{i+1}} $) have been generated by the same underlying stochastic process. We use a parametric spectral discrimination approach to solve this problem.

\subsection{Change points exploration by using scan statistics}
\label{sec:methods2}
In section \ref{sec:methods1}, we did not define the range of $q$, which could be any value between $0$ and $T$. Therefore, as the first step, a possibly overestimated set of change points will be estimated by using the likelihood ratio scan statistics proposed by \cite{yau2016inference}.
A brief introduction is given in this section.

For a window radius $h$ we define a corresponding scanning window $R_{t}(h)$ and observations as
\begin{align*}
 R_{t}(h) = {t-h+1, \dots, t+h}\\
 x_{R_{t}(h)} = {x_{t-h+1}, \dots, x_{t+h}}
\end{align*}
 The likelihood ratio scan statistics is then
\begin{align*}
 LS_{h}(t) & = \frac{1}{h}L_{t-h+1,\dots,t}(t,\hat{\theta}_{1}) + \frac{1}{h}L_{t+1,\dots,t+h}(t,\hat{\theta}_{2}) -
 \frac{2}{h}L_{t-h+1,\dots,t+h} (t,\hat{\theta}) ,\\
 & \text{where } L(\theta)  = \sum_{t=1}^{T}\log f_{\theta}(x_{t}\mid x_{t-1},\dots,x_{t-p})
\end{align*}
By scanning the observed time series data, a sequence of $LS_{h}(t)$ will be obtained at $t=$ $h$, $h+1$, \dots, $n-h$. If $h$ meets certain criteria, at most one change point outputs in $R_{t}(h)$, and if there is a change at $t$, then $LS_{h}(t)$ tends to be large. Hence, a set of potential change points $ \hat{k} =  (k_{1}, k_{2}, \dots, k_{q}) $ will be obtained after the scanning process.    

\subsection{A likelihood ratio test for comparing time series}
\label{sec:lrtest}
Given a set of estimated change points, we then apply a modified version of the parametric spectral discrimination test proposed by \cite{grant2017parametric} to test if the adjacent segments are from the same autoregressive process. We fit the autoregressive models%
%\[
\begin{eqnarray} 
x_{t}+\beta_{x,1}x_{t-1}+\ldots+\beta_{x,p_{x}}x_{t-j}=\varepsilon_{t} \nonumber \\
%\]
%\[
y_{t}+\beta_{y,1}y_{t-1}+\ldots+\beta_{y,p_{y}}y_{t-j}=u_{t},
%\]
\end{eqnarray}
to two adjacent segments of lengths $T_{1}$ and $T_{2}$, respectively, where $\left\{
\varepsilon_{t}\right\}  $ and $\left\{  u_{t}\right\}  $ are independent processes with zero mean and variances $\sigma_{\varepsilon}^{2}$ and $\sigma_{u}^{2}$, respectively. Although the test is developed as though $\left\{\varepsilon_{t}\right\}  $ and $\left\{  u_{t}\right\}  $ are i.i.d and Gaussian, the asymptotic distribution of the test statistic holds under much weaker conditions \citep{G2018}. Note that we are also assuming that the processes have zero mean, and in practice the time series are mean-corrected before analysis. That is, we do not consider a shift in mean between segments to constitute a change point, but rather consider only changes in the second-order properties. It is straightforward to adjust the test to consider a change in mean a change point. The hypothesis test is
\begin{align*}
H_{0}  &  : \beta_{X,j}=\beta_{Y,j}\quad\text{for all }j,\quad \sigma_{\varepsilon}^{2}=\sigma_{u}^{2}\\
H_{A}  &  :\text{Not }H_{0}\text{.}%
\end{align*}
Under the null hypothesis, the underlying processes share the same autocovariance structure, or, in other words, have the same spectral density (hence the term spectral discrimination tests). In order to compute the likelihood ratio statistic, we need the maximum likelihood estimators of the parameters under both $H_{0}$ and $H_{A}$. Under $H_{A}$, the processes are independent and the parameters can be estimated separately using, for example, the Levinson--Durbin algorithm \cite{Levinson1947, Durbin1960}. For a given order $p$, the algorithm computes the estimators%
%\[
\begin{align*}
\widehat{\beta}^{p} & =-\widehat{\Gamma}_{p}^{-1}\widehat{\gamma}^{p},\\
\widehat{\sigma}_{p}^{2} & =\widehat{\gamma}\left(  0\right)  +\left(
\widehat{\gamma}^{p}\right)  ^{\prime}\widehat{\beta}^{p},
%\]
\end{align*}
where%
\begin{align*}
%\[
\widehat{\beta}^{p} & =\left[
\begin{array}
[c]{ccc}%
\beta_{1} & \cdots & \beta_{p}%
\end{array}
\right]  ^{\prime}, \quad
%\]%
\widehat{\gamma}^{p}  =\left[
\begin{array}
[c]{ccc}%
\gamma\left(  1\right)  & \cdots & \gamma\left(  p\right)
\end{array}
\right]  ^{\prime}, \quad
\widehat{\gamma}\left(  j\right) =\frac{1}{T}\sum_{t=j}^{T-1}x_{t}x_{t-j},
\end{align*}
$T$ is the sample size and $\widehat{\Gamma}_{p}$ is the $p\times p$ matrix
with $\left( i,j\right)$th entry given by $\widehat{\gamma}\left(
\left\vert i-j\right\vert \right) $.
Under $H_{0}$, for $j=0,\ldots,p$, and given $\lambda$, we define
\[
c\left(  j\right)  =\frac{1}{T_{1}+T_{2}}\left(  \sum_{t=j}^{T_{1}-1}%
x_{t}x_{t-j}+\sum_{t=j}^{T_{2}-1}y_{t}y_{t-j}\right)  .
\]
Replacing $\widehat{\gamma}\left(  j\right)  $ by $c\left(  j\right)  $ in the
Levinson--Durbin algorithm gives estimators for the common parameters. The test statistic is%
\[
\Lambda=T_{1}\log\left(  \frac{\widehat{\sigma}_{0}^{2}%
}{\widehat{\sigma}_{\varepsilon;A}^{2}}\right)  +T_{2}\log\left(
\frac{\widehat{\sigma}_{0}^{2}}{\widehat{\sigma}_{u;A}^{2}}\right)  ,
\]
where $\widehat{\sigma}_{\varepsilon;A}^{2}$ and $\widehat{\sigma}_{u;A}^{2}$
are the estimators of $\sigma_{\varepsilon}^{2}$ and $\sigma_{u}^{2}$ under
$H_{A}$, and $\widehat{\sigma}_{0}^{2}$ is the estimator of the common residual variance under $H_{0}$. We reject $H_{0}$ when $\Lambda$ is greater than the
$100\left(  1-\alpha\right)  $th percentile of the $\chi^{2}$ distribution
with $p_{x}+p_{y}-p+1$ degrees of freedom.

Since the orders are unknown in practice, they can be estimated using, for
example, an information criterion such as BIC. This is easily incorporated
into the Levinson--Durbin algorithm. However, it was shown in \cite{grant2017parametric} that the test
performs poorly when the underlying time series are not truly autoregressive. The proposed solution was to use autoregressive approximation by fixing the orders, under both $H_{0}$ and $H_{A}$, as $p_{x}=p_{y}=p=\left\lfloor \left( \log T_{\min}\right)  ^{v}\right\rfloor $, where
$v>1$, $T_{\min}=\min\left(  T_{1},T_{2}\right)  $ and $\left\lfloor
\left( \log T_{\min}\right)  ^{v} \right\rfloor $ is the integer component of $\left( \log T_{\min}\right)  ^{v} $. The null hypothesis is then rejected when $\Lambda$ is greater than the $100\left(  1-\alpha\right)  $th percentile of the $\chi^{2}$ distribution
with $p+1$ degrees of freedom.  The test then performs well even when the time series are not autoregressive, with the cost being some loss in power in the autoregressive case.

\subsection{Approaches for multiple hypothesis tests}
Generally, for a single hypothesis test, we specify a Type I error, say 0.05, and make a conclusion based on the test statistic which meets this specification while giving the highest power. When multiple hypotheses are tested simultaneously, the probability of at least one incorrect ``statistically significant" outcome is increased with as the number of independent tests increases, which may result in incorrect conclusions. Thus, it is necessary to evaluate the tests as a whole. Numerous procedures have been proposed for this multiple comparison problem. In this paper, we implement two classical procedures: Controlling the false discovery rate, proposed by \cite{benjamini1995controlling} (BH); and the adjusted p-values approach of \cite{wright1992adjusted}.     

As per the previous subsection, we can obtain unadjusted p-values $p_{(1)},p_{(2)}$, $\dots,$ $p_{(q)}$ corresponding to the multiple hypotheses considered in (\ref{2.3}). Let $P_{(1)} \leq P_{(2)}\leq 
\dots \leq P_{(q)}$ be the ordered $p_{(1)},p_{(2)}, \dots,p_{(q)}$ from smallest to largest. The BH multiple-testing procedure is as follows.
\begin{equation}
\text{For each $i =1,2,\dots,q$, if } P_{(i)} \leq \frac{i}{q}\alpha \nonumber \\
\end{equation} 
\begin{equation}
\text{then reject all }H_{(i)}\nonumber\\
\end{equation}
\begin{center}
\mbox{$ \hat{k}^{*} = (k_{1}, k_{1}, \dots, k_{q^{*}})$ is the final estimates of change points.} \nonumber 
\end{center}
Next, we adopt the adjusted p-values method by Bonferroni procedure as follows.
\begin{equation}
\text{For each $i =1,2,\dots,q$, if }q \times p_{(i)} \leq \alpha \nonumber \\
\end{equation} 
\begin{equation}
\text{then reject all } H_{(i)} \nonumber
\end{equation} 
\begin{center}
\mbox{$ \hat{k}^{*} = (k_{1}, k_{1}, \dots, k_{q^{*}})$ is the final estimates of change points.} \nonumber 
\end{center}

\section{Simulation Study}
\label{sec:sims}
In this section, we use nine classic examples to compare the performance of the MCP2 method with methods from recent literature including the likelihood ratio scan method (LRSM) by \cite{yau2016inference} and the wild binary segmentation method (WBS) by \cite{korkas2017multiple}. Except model G, the models that are used here were also considered by \cite{yau2016inference}, hence, we apply the same settings. For each model, we simulated 100 sequences. In addition, we used two simple measurements to evaluate the estimation accuracy: the estimated number of change points and the estimated locations. Table \ref{3.1} provides the exact detection rate for each model, which describes the ability for each method to estimate the correct number of change points among 100 sequences. It is then straightforward to get insights into the distance between the true and estimated locations of change points through the use of novel plots, which provide a good visualisation of estimation accuracy for different methods.   

\begin{table}[ht] \centering
\begin{tabular}
[c]{llllll}\hline
& & \textbf{MCP2BH} & \textbf{MCP2WRI} & \textbf{LRSM} & \textbf{WBS}\\\hline
Model A&$\beta =  -0.7 $ & 0.77 & 0.79 & 1 & 0.35 \\
&$\beta =  -0.1$ & 0.66 & 0.69 & 1 & 0.95 \\
&$\beta =  0.4$ & 0.67 & 0.68 & 1  & 0.93 \\
&$\beta =  0.7$ & 0.68 & 0.69 & 1 & 0.93 \\ \hline
Model B& & 0.61& 0.76 & 0.95 & 0.52 \\ 
Model C& & 0.80& 0.94 & 0.99 & 0.88 \\ 
Model D& & 0.68& 0.81 & 0.97 & 0.67 \\ 
Model E& & 0.32& 0.35 & 0.21 & 0.22 \\ 
Model F& & 0.34& 0.36  & 0.23 & 0.32 \\ 
Model G& & 0.70& 0.76 & 0.38 & 0.38 \\ 
Model H& & 0.73& 0.82 & 0.55 & 0.54 \\ 
Model I& & 1& 1 & 1 & 0.97 \\ \hline
\end{tabular}
\caption{Exact detection rate of estimated number of change points for each method. The rate is the number of correct estimation over 100.}
\label{3.1}
\end{table}%

\begin{enumerate}
\item[a] Model A: stationary AR(1) process with various $\beta = -0.7, -0.1, 0.4, 0.7$
\begin{equation}
x_{t}= \beta x_{t-1} + \varepsilon_{t}, 1 \leq t \leq 1024
\end{equation} 
We evaluate the performance of the methods via Model A that there is no change point. LRSM is overall perfect under model A, WBS is nearly perfect except the poor performance when $\beta = -0.7$. MCP2 method performs well and almost uniformly with various $\beta$.       

\item[b] Model B: piecewise stationary auto-regressive process 
\begin{equation} 
x_{t}=
\begin{cases}
0.9x_{t-1} + \varepsilon_{t} &1 \leq t \leq 512\\
1.69x_{t-1} - 0.81x_{t-2} + \varepsilon_{t}, &513 \leq t \leq 768\\
1.32x_{t-1} - 0.81x_{t-2} + \varepsilon_{t}, &769 \leq t \leq 1024\\
\end{cases}
\end{equation}

\begin{figure}[ht]
	\center{\includegraphics[width=\textwidth]{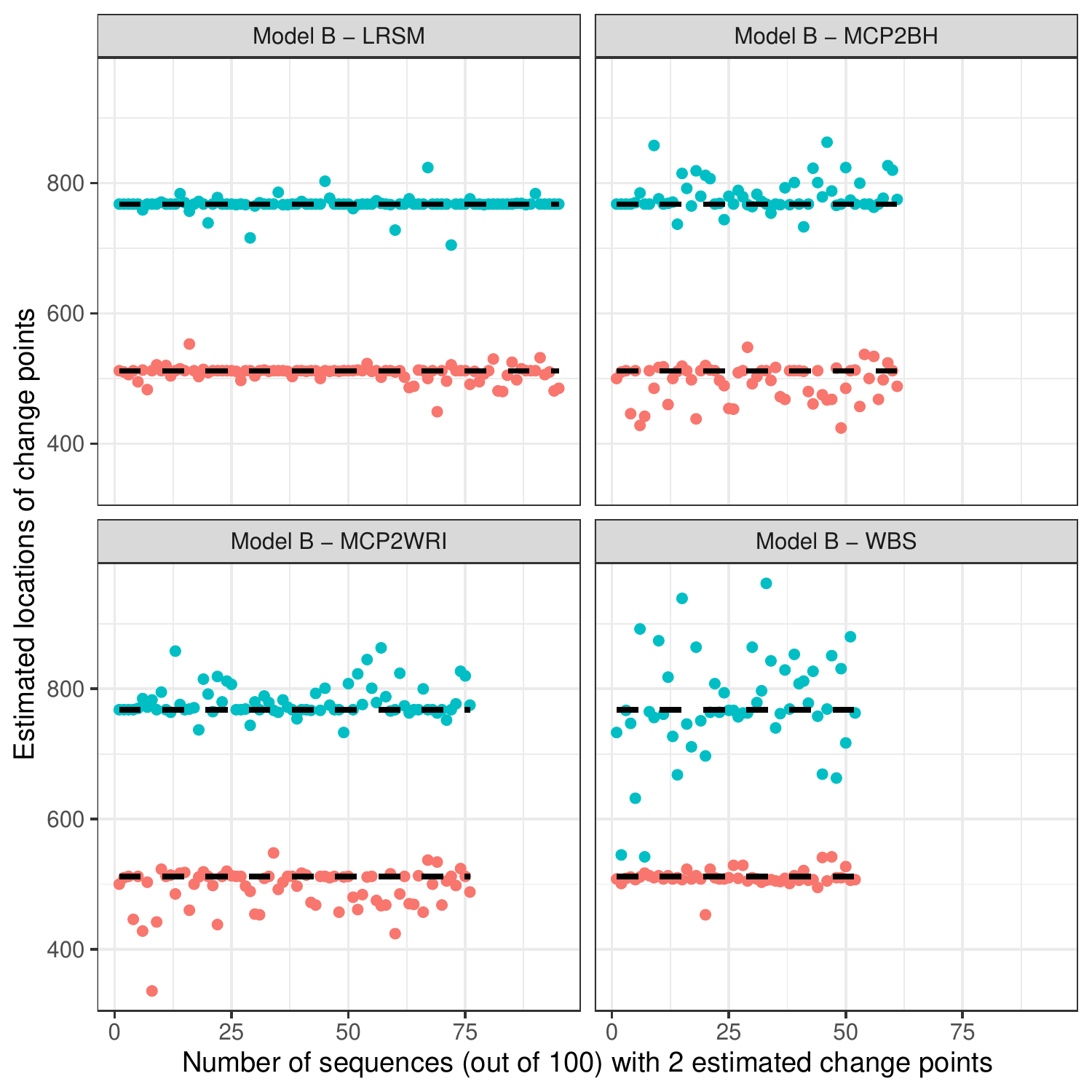}}
	\caption{Plots of estimated locations of change points from different methods under model B. Horizontal line stands for the sequence of estimated changes only when the estimated number of change points equals to 2. The dashed black lines represent the true locations of change points, 512 and 768.}
	\label{3.2}
\end{figure}

In terms of the performance of estimating number of change points, LRSM is outstanding over the others, and WBS has the lowest accuracy rate. Moreover, LRSM gives the most accurate estimated locations which can be seen by looking at Figure \ref{3.2}. WBS seems to lose power at time point 768. The performance of MCP2 is in between.

\item[c] Model C: piecewise stationary AR(1) process 
\begin{equation} 
x_{t}=
\begin{cases}
0.4x_{t-1} + \varepsilon_{t} &1 \leq t \leq 400\\
-0.6x_{t-1} + \varepsilon_{t}, &401 \leq t \leq 612\\
0.5x_{t-1} + \varepsilon_{t}, &613 \leq t \leq 1024\\
\end{cases}
\end{equation} 

\begin{figure}[ht]
	\center{\includegraphics[width=\textwidth]{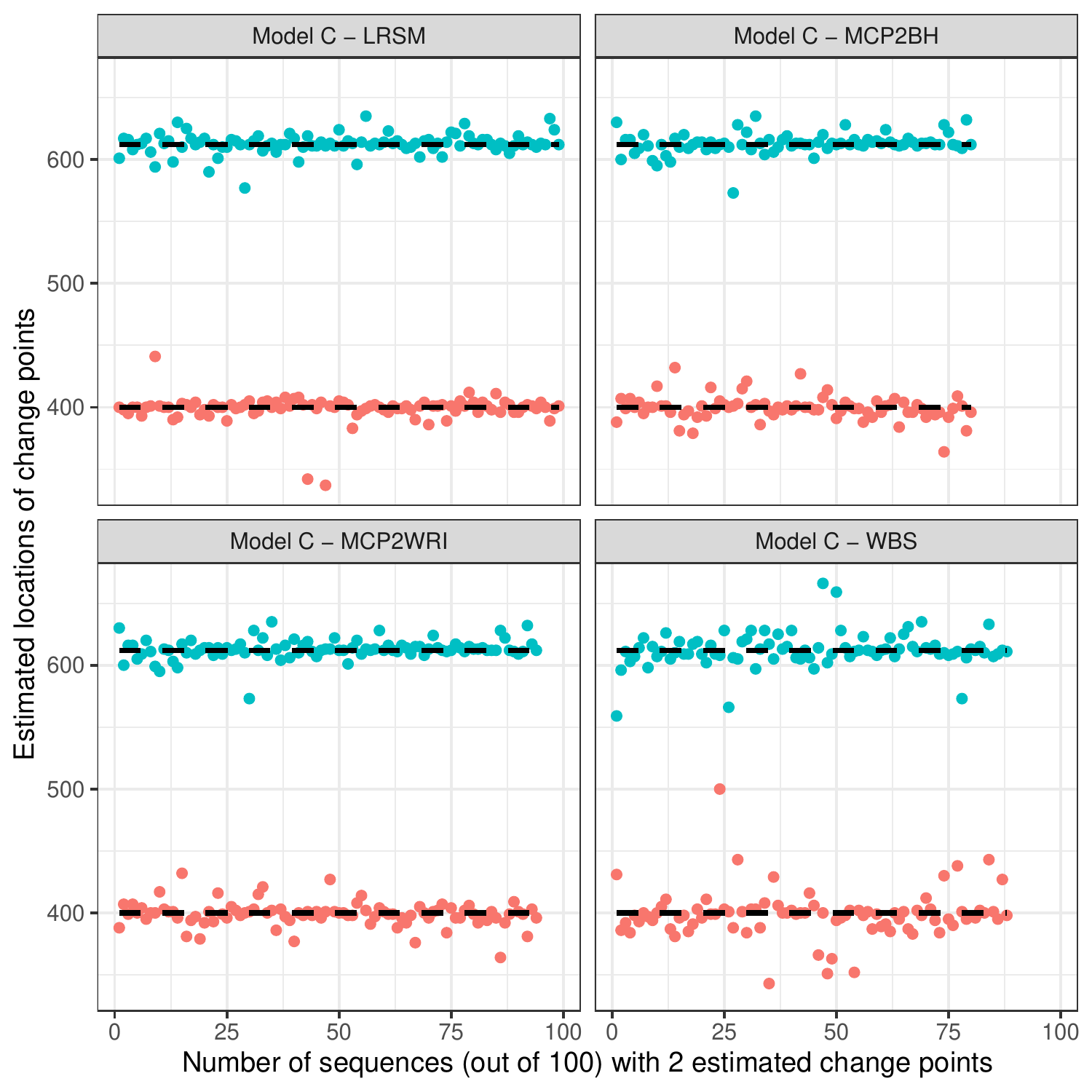}}
	\caption{Plots of estimated locations of change points from different methods under model B. Horizontal line stands for the sequence of estimated changes only when the estimated number of change points equals to 2. The dashed black line represents the true locations of change points, 400 and 750.}
	\label{3.3}
\end{figure}

Comparing with model B, the performance of all methods improved for both the estimates of number and locations of change points. Again, the LRSM is superior over MCP2 and WBS. 

\item[d] Model D: piecewise stationary AR(1) process with a short segment  
\begin{equation} 
x_{t}=
\begin{cases}
0.75x_{t-1} + \varepsilon_{t} &1 \leq t \leq 50\\
-0.5x_{t-1} + \varepsilon_{t}, &51 \leq t \leq 1024\\
\end{cases}
\end{equation} 

\begin{figure}[ht]
	\center{\includegraphics[width=\textwidth]{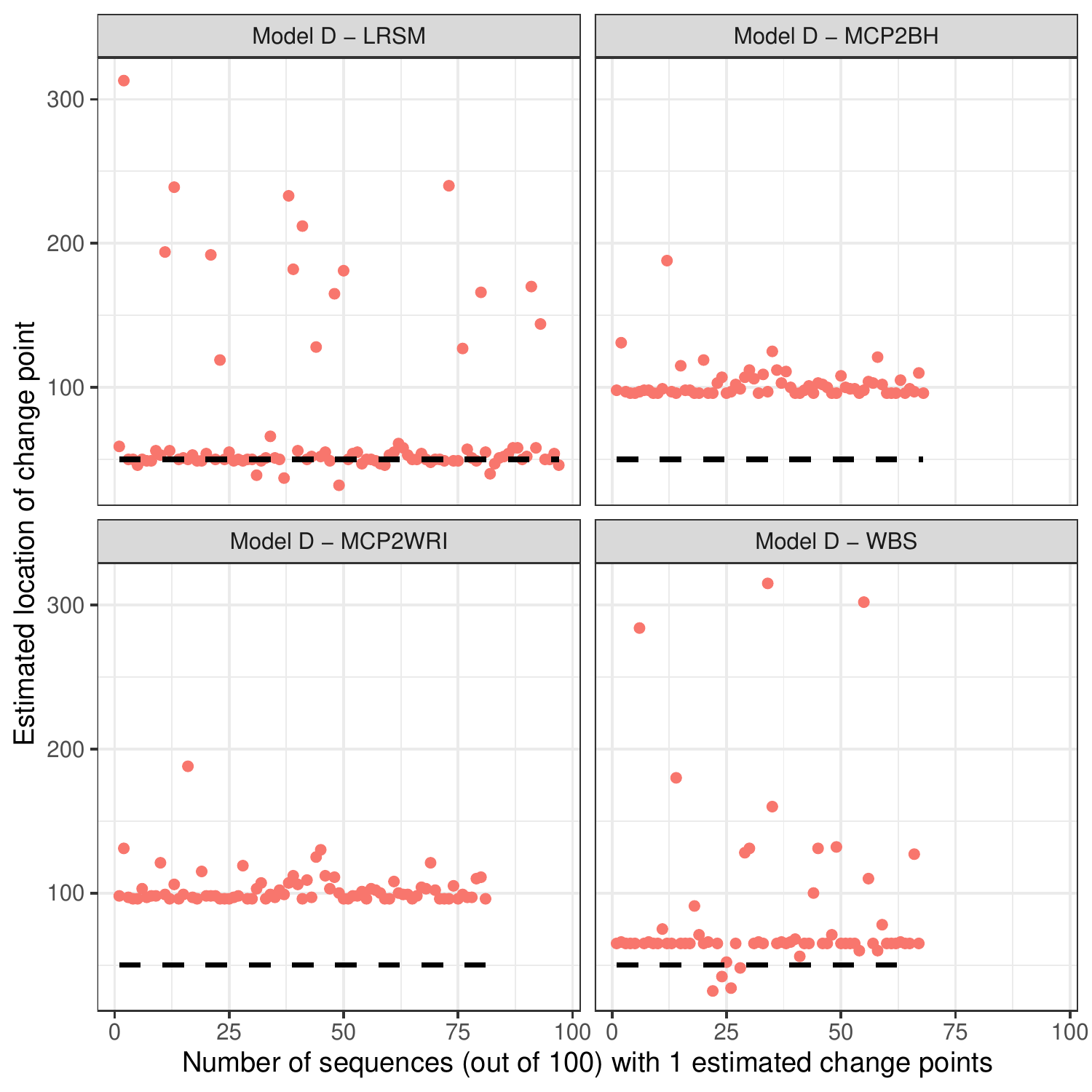}}
	\caption{Plots of estimated locations of change points from different methods under model B. Horizontal line stands for the sequence of estimated changes only when the estimated number of change points equals to 1. The dashed black line represents the true location of change points at 50.}
	\label{3.4}
\end{figure}
LRSM remains the outstanding method in estimating the number of change points and locations compared with the others. It is obvious that the MCP2 method failed to detect the change point located at time 50, which is toward the very beginning of the series, as shown in Figure \ref{3.4}. WBS is in between the other two.  

\item[e] Model E: piecewise stationary near-unit-root process with changing variance 
\begin{equation} 
x_{t}=
\begin{cases}
0.999x_{t-1} + \varepsilon_{t} &\varepsilon_{t}\sim N(0,1), 1 \leq t \leq 400\\
0.999x_{t-1} + \varepsilon_{t}, &\varepsilon_{t}\sim N(0,1.5^2),401 \leq t \leq 750\\
0.999x_{t-1} + \varepsilon_{t}, &\varepsilon_{t}\sim N(0,1),751 \leq t \leq 1024\\
\end{cases}
\end{equation} 
Since the autocorrelation coefficients of this series remain unchanged for each segment and close to 1, all methods do not perform well.  
\item[f] Model F: piecewise stationary AR process with high autocorrelation
\begin{equation} 
x_{t}=
\begin{cases}
1.399x_{t-1} - 0.4x_{t-2} + \varepsilon_{t} &\varepsilon_{t}\sim N(0,1), 1 \leq t \leq 400\\
0.999x_{t-1} + \varepsilon_{t}, &\varepsilon_{t}\sim N(0,1.5^2),401 \leq t \leq 750\\
0.699x_{t-1} + 0.3x_{t-2} + \varepsilon_{t}, &\varepsilon_{t}\sim N(0,1),751 \leq t \leq 1024\\
\end{cases}
\end{equation} 

Simulations from models E and F are challenging data sets. From the table \ref{3.1}, the detection rate for all methods is quite low at around 0.3. Hence, it is not useful to plot the corresponding locations. MCP2 performed slightly better than the other two methods.

\item[g] Model G: piecewise stationary AR(1) process with three change points
\begin{equation} 
x_{t}=
\begin{cases}
0.7x_{t-1} + \varepsilon_{t} &1 \leq t \leq 125\\
0.3x_{t-1} + \varepsilon_{t} &126 \leq t \leq 532\\
0.9x_{t-1} + \varepsilon_{t} &533 \leq t \leq 704\\
0.1x_{t-1} + \varepsilon_{t} &705 \leq t \leq 1024\\
\end{cases}
\end{equation} 

\begin{figure}[ht]
	\center{\includegraphics[width=\textwidth]{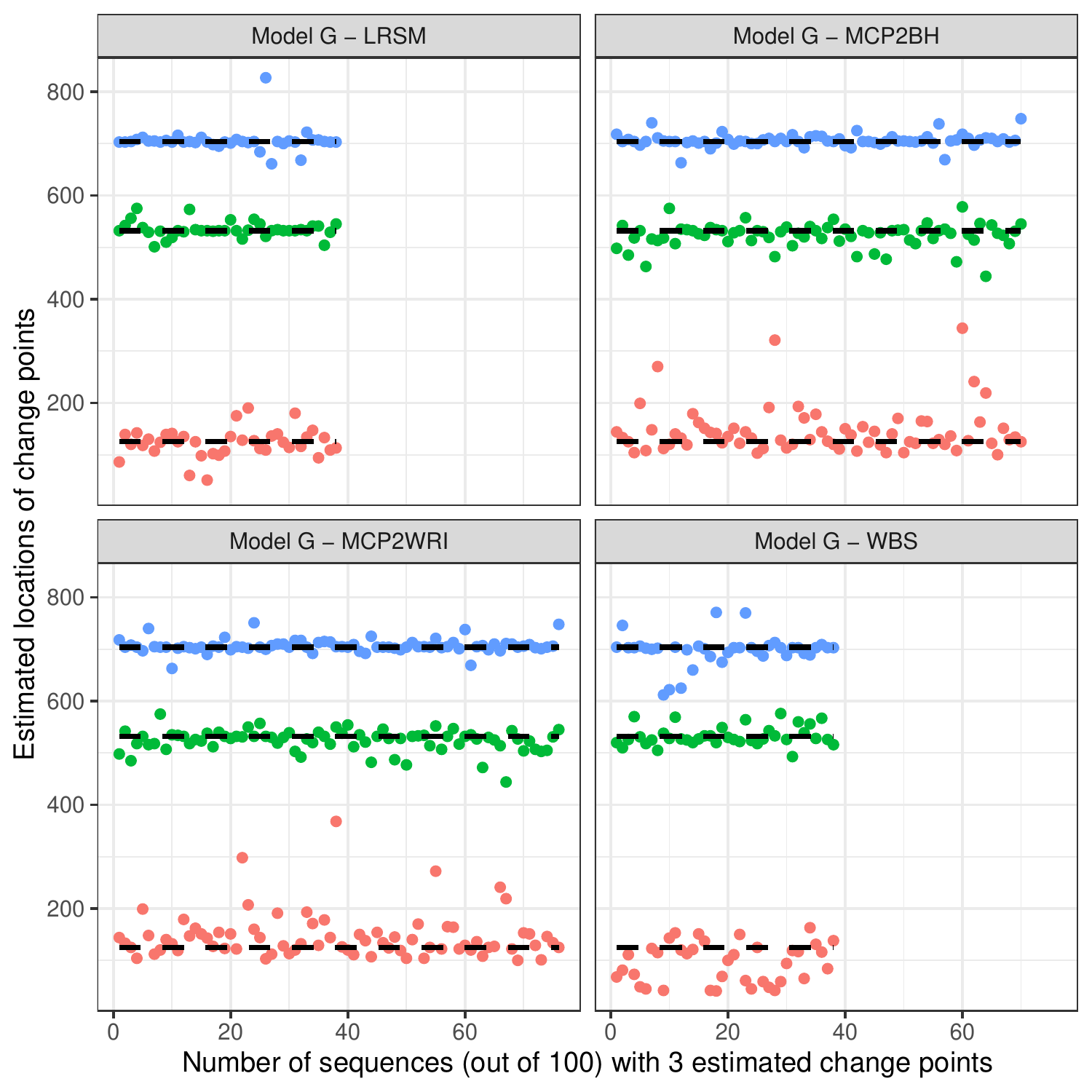}}
	\caption{Plots of estimated locations of change points from different methods under model B. Horizontal line stands for the sequence of estimated changes only when the estimated number of change points equals to 3. The dashed black line represents the true locations of change points, 125, 532 and 704.}
	\label{3.5}
\end{figure}
MCP2 outperformed the other methods under this model in terms of estimating the number of change points. WBS and LRSM had similar performance.    

\item[h] Model H: piecewise stationary ARMA(1,1) process with three change points
\begin{equation} 
x_{t}=
\begin{cases}
0.7x_{t-1} + \varepsilon_{t} + 0.6\varepsilon_{t-1} &1 \leq t \leq 125\\
0.3x_{t-1} + \varepsilon_{t} + 0.3\varepsilon_{t-1} &126 \leq t \leq 532\\
0.9x_{t-1} + \varepsilon_{t}                        &533 \leq t \leq 704\\
0.1x_{t-1} + \varepsilon_{t} - 0.5\varepsilon_{t-1} &705 \leq t \leq 1024\\
\end{cases}
\end{equation} 

\begin{figure}[ht]
	\center{\includegraphics[width=\textwidth]{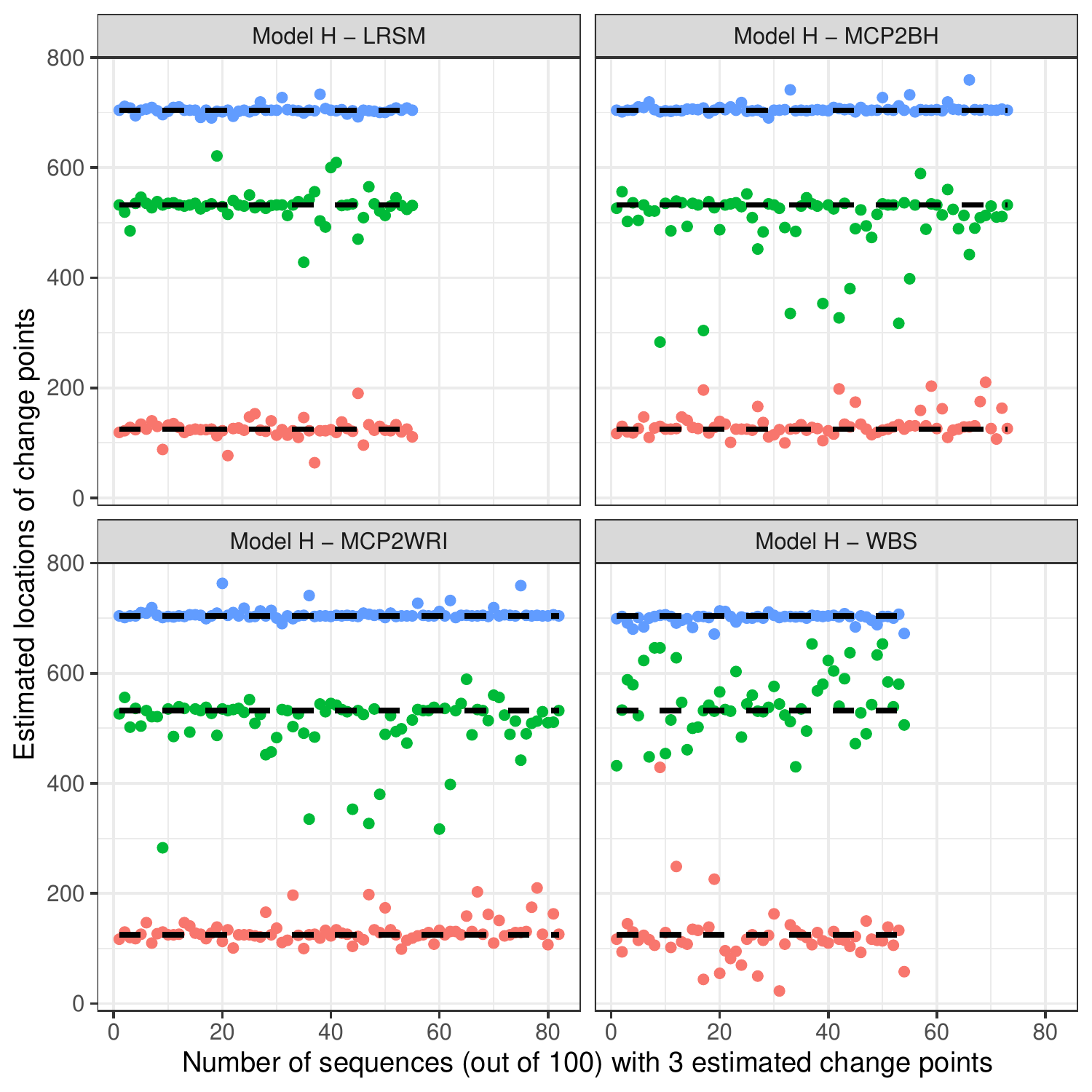}}
	\caption{Plots of estimated locations of change points from different methods under model B. Horizontal line stands for the sequence of estimated changes only when the estimated number of change points equals to 3. The dashed black line represents the true locations of change points, 125, 532 and 704.}
	\label{3.6}
\end{figure}
Similar to the previous model, MCP2 had the best performance when estimating the number of change points, while it is interesting to see that WBS and MCP2 struggled to detect the second change point. Comparing WBS with LRSM, LRSM remains robust when estimating the locations.

\item[i] Model I: piecewise stationary moving average process
\begin{equation} 
x_{t}=
\begin{cases}
\varepsilon_{t} + 0.8\varepsilon_{t-1} &1 \leq t \leq 128\\
\varepsilon_{t} + 1.68\varepsilon_{t-1} - 0.81\varepsilon_{t-2}  &129 \leq t \leq 256\\
\end{cases}
\end{equation} 

\begin{figure}[ht]
	\center{\includegraphics[width=\textwidth]{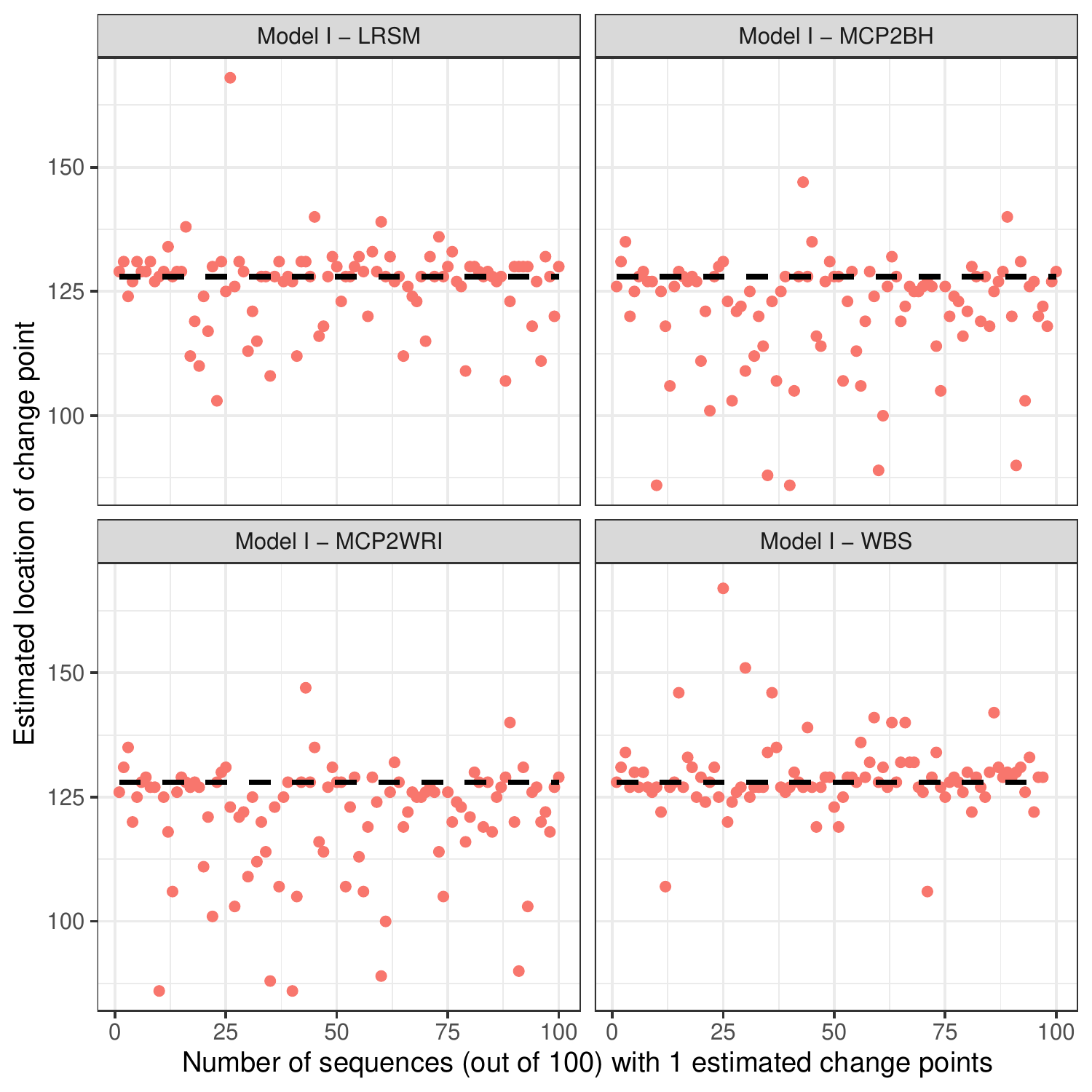}}
	\caption{Plots of estimated locations of change points from different methods under model B. Horizontal line stands for the sequence of estimated changes only when the estimated number of change points equals to 1. The dashed black line represents the true location of change point at 128.}
	\label{3.7}
\end{figure}
All methods performed perfect when estimating the number of change points, while in terms of estimating the locations, all methods perform poorly.

\end{enumerate}

\section{Real Data Analysis}
\label{sec:data}

\subsection{Example 1: Physiological data time series}
\begin{figure}[ht]
	\center{\includegraphics[width=\textwidth]{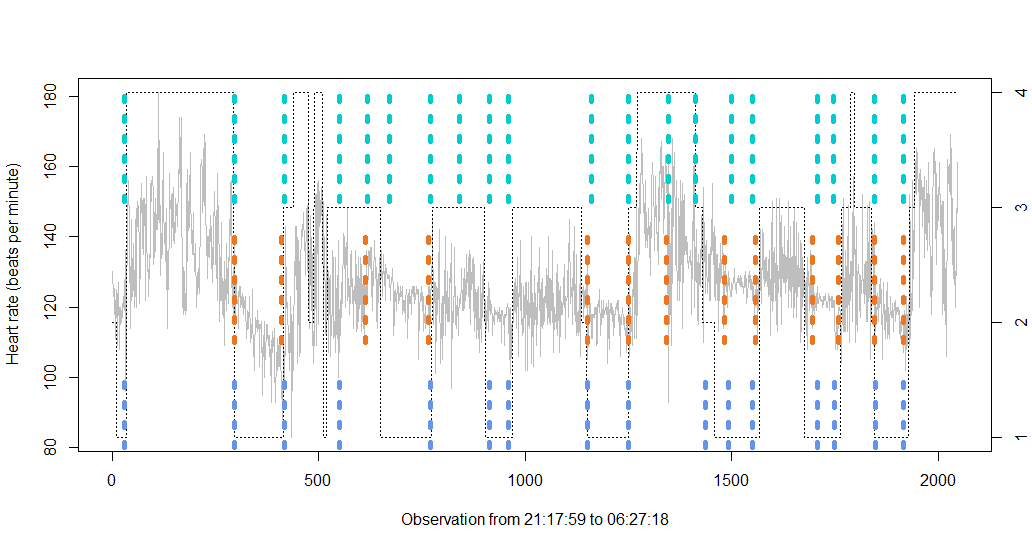}}
	\caption{Performance of MCP2 with WBS, the top and bottom dotted line represents MCP2-BH and MCP2-Wright, the middle dotted line represents WBS method with default setting. The right hand axis represents 1=quiet sleep, 2=between quiet and active sleep, 3=active sleep, 4=awake.}
	\label{4.1}
\end{figure}

In this section, we use two linked medical time series, BabyECG and BabySS, which are available in the R package \textit{wavethresh}, containing 2048 observations of an infant's heart rate and sleep state sampled every 16 seconds recorded from 21:17:59 to 06:27:18. Both of them were recorded from the same 66 day old infant. The dashed line represents a change in sleep state. \cite{korkas2017multiple} has analysed the BabyECG time series as a real data example of a piecewise stationary time series by using the WBS method. Here we compare MCP2 with WBS, since LRSM is not applicable for this situation.
From Figure \ref{4.1}, it can be seen that all methods tend to be in agreement at most estimated change points. MCP2 is able to identify the short segment if we use the smallest scanning window whereas WBS may ignore the shorter segments. In addition, the BH procedure is more conservative than Wright's. We remark that the selection of a scanning window exerts a control on the final estimates. In this situation, the scanning window we use is $h = \max \left\lbrace  50,\log(2048) \right\rbrace $.        

\subsection{Example 2: Monthly IBM stock returns}
The experiment we perform here is used for comparing MCP2 with the LRSM method by analysing monthly stock returns of IBM from January 1962 to October 2014, which is an example tested by \cite{yau2016inference} using LRSM. The scanning window used in MCP2 is the same as LRSM, which is $h=41$. LRSM gives two changes at 307 (July 1987) and 491 (November 2002), whereas MCP2-BH gives two estimations at 390 (June 1994) and 492 (December 2002). MCP2-Wright gives only one detection at 492. It seems that there is a clear agreement on the second change point.   
\begin{figure}[ht]
	\center{\includegraphics[width=\textwidth]{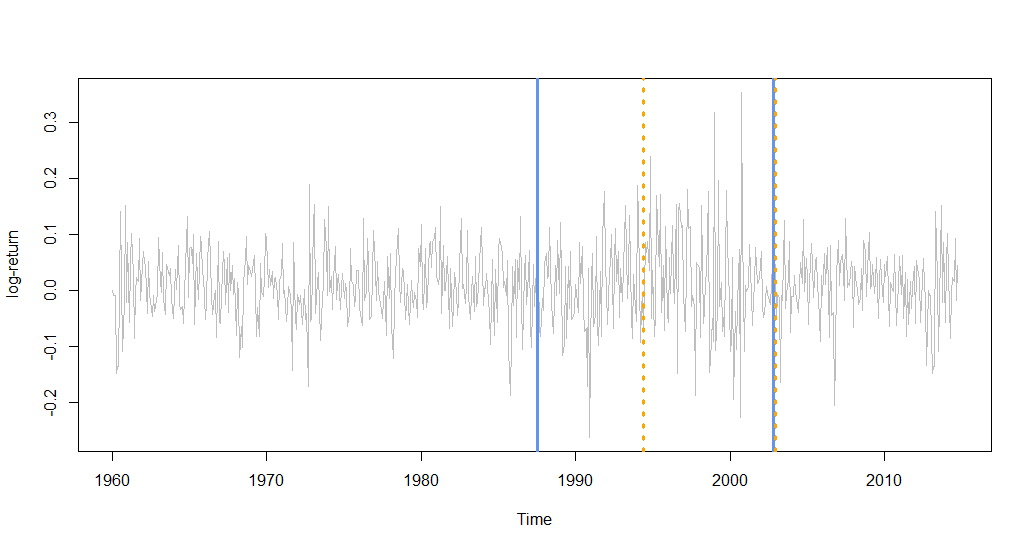}}
	\caption{Performance of MCP2 with LRSM, the blue line represents LRSM, the orange dotted line represents MCP2-BH method.}
	\label{4.2}
\end{figure}

\section{Conclusion}
\label{sec:discussion}
In this paper, we proposed the MCP2 method which shows the flexibility and superior performance over the LRSM and WBS methods in piecewise stationary autoregressive process with more than two change points. In terms of measuring the locations of change points, we used novel statistical plots instead of the Hausdorff distance, the advantage being that we can get insights from the plots as to what caused the over-segmentation. In addition, the plots clearly demonstrated the performance of each method when estimating the locations of change points. Future work will involve a theoretical investigation of our method as well as work to further improve the estimation accuracy.

%\bibliographystyle{plainnat}  


\end{document}